\documentclass[aps,prd,eqsecnum,showpacs,amsmath,amssymb,twocolumn,superscriptaddress,nofootinbib]{revtex4-1}

\usepackage{graphicx}
\usepackage{dcolumn}
\usepackage{bm}
\usepackage{color}
\usepackage{subfigure}

\newcommand{\be}{\begin{equation}}
\newcommand{\ee}{\end{equation}}
\newcommand{\bea}{\begin{eqnarray}}
\newcommand{\eea}{\end{eqnarray}}
\newcommand{\non}{\nonumber}
\newcommand{\Non}{\nonumber\\}
\newcommand{\bseq}{\begin{subequations}}
\newcommand{\eseq}{\end{subequations}}

\allowdisplaybreaks[1]
\begin{document}

%\vspace{0.4cm}

%%%%%%%%%%%%%%%%%%%%%%%%%%%%%%%%%%%%%%%%%%%%%%%%%%%%%%%%%%
\title{Kinetic gravity braiding wormhole geometries}
%%%%%%%%%%%%%%%%%%%%%%%%%%%%%%%%%%%%%%%%%%%%%%%%%%%%%%%%%%

\author{Roman Korolev}
\email{korolyovrv@gmail.com}
\affiliation{Institute of Physics, Kazan Federal University, Kremliovskaya str. 16a, Kazan 420008, Russia}

\author{Francisco S. N. Lobo}
\email{fslobo@fc.ul.pt}
\affiliation{Departamento de F\'{i}sica and Instituto de Astrof\'{i}sica e Ci\^{e}ncias do Espa\c{c}o, Faculdade de Ci\^encias da Universidade de Lisboa, Edif\'{i}cio C8, Campo Grande, P-1749-016, Lisbon, Portugal}
 
\author{Sergey V. Sushkov}
\email{sergey$_\,$sushkov@mail.ru}
\affiliation{Institute of Physics, Kazan Federal University, Kremliovskaya str. 16a, Kazan 420008, Russia}

\date{\today}

%%%%%%%%%%%%%%%%%%%%%%%%%%%%%%%%%%%%%%%%%%%%%%%%%%%%%%%%%%
\begin{abstract}
%%%%%%%%%%%%%%%%%%%%%%%%%%%%%%%%%%%%%%%%%%%%%%%%%%%%%%%%%%
	An interesting class of scalar-tensor models, denoted by kinetic gravity braiding (KGB), has recently been proposed. These models contain interactions of the second derivatives of the scalar field that do not lead to additional degrees of freedom and exhibit peculiar features, such as an essential mixing of the scalar $\phi$ and tensor kinetic $X$ terms. In this work, we consider the possibility that wormhole geometries are sustained by the KGB theory. More specifically, we present the full gravitational field equations in a static and spherically symmetric traversable wormhole background, and outline the general constraints at the wormhole throat, imposed by the flaring-out conditions. Furthermore, we present a plethora of analytical and numerical wormhole solutions by considering particular choices of the KGB factors. The analysis explicitly demonstrates that the KGB theory exhibits a rich structure of wormhole geometries, ranging from asymptotically flat solutions to asymptotically anti-de Sitter spacetimes.

%{\sf *** WRITE OUT AND FINISH OFF AS THE PAPER DEVELOPS ***}

\end{abstract}

\maketitle

%%%%%%%%%%%%%%%%%%%%%%%%%%%%%%%%%%%%%%%%%%%%%%%%%%%%%%%%%%
\section{Introduction}
%%%%%%%%%%%%%%%%%%%%%%%%%%%%%%%%%%%%%%%%%%%%%%%%%%%%%%%%%%

In modern physics, scalar fields are extremely appealing as they are ubiquitous in theories of high energy physics beyond the standard model and are popular building blocks used to construct cosmological models, such as in the late-time cosmic speed-up \cite{Perlmutter:1998np,Riess:1998cb}. In the latter cosmological context, the initial simplest model consisted of a minimally coupled single scalar field self-interacting through a scalar potential, and with canonical kinetic terms, denoted by ``quintessence'' \cite{Zlatev:1998tr}. An interesting model, described as coupled quintessence, was also put forward that analysed the consequences of coupling the field responsible for acceleration to matter fields \cite{Amendola:1999er}. In fact, a large number of models presently exist, and we refer the reader to \cite{Copeland:2006wr} for more details. 

Taking into account the models that abound in the literature, one may wonder how one should study and compare them in a unified manner, and determine which if any is the origin of cosmic acceleration. A particularly useful tool in this direction is the realisation that all these classes of models are special cases of the most general Lagrangian which leads to second order field equations, namely, the Horndeski Lagrangian \cite{Horndeski:1974wa}, which was recently rediscovered \cite{Deffayet:2011gz}. The Horndeski theory enables researchers to adopt a unifying framework, and to determine the regions within this general theory that have appealing theoretical properties. 
In addition to the need to adequately fit observations, these properties are useful in preferring specific regions of the general theory, which is equivalent to choosing particular models (see \cite{Kase:2018aps} for a recent review). 

In this context, an interesting class of scalar-tensor models containing interactions of the second derivatives of the scalar field, that do not lead to additional degrees of freedom, has been presented. This class of models exhibits peculiar features, such as an essential mixing of scalar and tensor kinetic terms \cite{Deffayet:2010qz}, and has been denoted by kinetic gravity braiding (KGB). It has been noted that the braiding causes the scalar stress tensor to deviate from the perfect-fluid form \cite{Pujolas:2011he}, and in particular, the cosmological models possess a rich phenomenology. It was found that the late-time asymptotic is a de Sitter state, and the scalar field can exhibit a phantom behaviour that is able to cross the phantom divide with neither ghosts nor gradient instabilities.

The study on the growth history of the linear density perturbation as well as the spherical collapse in the nonlinear regime of the density perturbations has also been undertaken, which is important in order to distinguish between the kinetic braiding and the $\Lambda$CDM (Lambda cold dark matter) models \cite{Kimura:2010di}.
Furthermore, the observational constraints on KGB from the ISW-LSS (Integrated Sach-Wolfe--Large Scale Structures) cross-correlation has also been investigated \cite{Kimura:2011td}. Indeed, it was found that the late-time ISW effect in the KGB model anti-correlates with large scale structures in a wide range of parameters, which clearly demonstrates how one can distinguish modified gravity theories from the $\Lambda$CDM model using the ISW effect.

A new class of inflationary models was also recently presented, consisting of a novel mechanism originating from a higher derivative KGB of an axion field \cite{Maity:2012dx}. It was found that a huge parameter regime exists in the model where the axion decay constant could be naturally sub-Planckian. This effectively reduced Planck scale provides one with the mechanism of further lowering the speed of an axion field rolling down its potential without introducing a super-Planckian axion decay constant.
Within the KGB class of models, the most general form of the Lagrangian that allows for cosmological scaling solutions was also found \cite{Gomes:2013ema}. It was argued that these scaling solutions may help in solving the coincidence problem, since in this case the presently observed ratio of matter to dark energy does not depend on the initial conditions, but rather on the theoretical parameters.
In the context of the two-field measure theory, it was also shown that KGB models can lead to phantom dark energy, stiff matter, and a cosmological constant behaviours \cite{Cordero:2019mze}.

Recently, astrophysical compact objects such as hairy stealth black holes \cite{Bernardo:2019yxp} were explored and the perturbations were further studied \cite{Bernardo:2020ehy} in shift symmetric KGB. Static and spherically symmetric wormholes geometries at the throat have also been analysed in the KGB theory \cite{Korolev:2020ohi}. In fact, in a recent work \cite{Korolev:2020ohi}, we considered the full Horndeski Lagrangian applied to wormhole geometries and presented the full gravitational field equations. We analysed the general constraints imposed by the flaring-out conditions at the wormhole throat and considered a plethora of specific subclasses of the Horndeski Lagrangian, namely, quintessence/phantom fields, $k$-essence, scalar-tensor theories, covariant galileons, nonminimal kinetic coupling, kinetic gravity braiding, and the scalar-tensor representation of Gauss-Bonnet couplings, amongst others. We argued that the generic constraints analysed served essentially as a consistency check of the main solutions obtained in the literature and drew out new avenues of research in considering applications of specific subclasses of the Horndeski theory to wormhole physics.

In the present work, we build on this analysis, in particular from the conditions explored in \cite{Korolev:2020ohi}, and consider specific wormhole solutions of the KGB theory.
In fact, wormhole solutions in modified theories of gravity are particularly interesting, as in General Relativity these exotic geometries \cite{Morris:1988cz,Morris:1988tu,Visser:1995cc,Lobo:2017oab} violate the energy conditions \cite{Sushkov:2005kj,Lobo:2005us,Lobo:2005yv,Capozziello:2013vna,Capozziello:2014bqa,Mimoso:2014ofa,Lobo:2009ip}. Now, it was shown in modified gravity that one may impose that the matter threading the wormhole satisfies the energy conditions, so that it is the effective energy-momentum tensor containing higher order curvature derivatives that is responsible for the null energy condition violation \cite{Harko:2013yb}. Thus, the higher order curvature terms, interpreted as a gravitational fluid, essentially sustain these non-standard wormhole geometries, that are fundamentally different from their counterparts in GR \cite{Lobo:2009ip,Bohmer:2011si,Sushkov:2011jh,Korolev:2014hwa,Bronnikov:2015aam,Harko:2013yb,Lobo:2007qi,MontelongoGarcia:2010xd,Garcia:2010xb,Lobo:2008zu,Mehdizadeh:2015jra,Boehmer:2007rm,Zangeneh:2015jda,Lobo:2010sb,Harko:2013aya,Boehmer:2007rm,Bronnikov:2010tt,Bronnikov:2002sf,Bronnikov:2006pt,Bronnikov:2012ch,Bronnikov:2019ugl}.

This paper is organised in the following manner: In Sec. \ref{sec:formalism}, we briefly present the formalism of the KGB theory and deduce the field equations in a static and spherically symmetric wormhole background, that are to be used throughout the analysis. In Sec. \ref{sec:constraintthroat}, we outline the general constraints at the wormhole throat, imposed by the flaring-out conditions, and consider specific cases. In Sec. \ref{sec:specificsolution}, we present specific wormhole solutions in the KGB theory, and demonstrate the existence of a large class of solutions, ranging from asymptotically flat solutions to asymptotically anti-de Sitter spacetimes. Finally, in Sec. \ref{sec:conclusion}, we conclude and discuss our results.

%{\color{blue}{\sf *** IMPROVE AS THE PAPER DEVELOPS ***} }

%%%%%%%%%%%%%%%%%%%%%%%%%%%%%%%%%%%%%%%%%%%%%%%%%%%%%%%%%%
\section{Wormhole configuration in the kinetic gravity braiding theory}\label{sec:formalism}
%%%%%%%%%%%%%%%%%%%%%%%%%%%%%%%%%%%%%%%%%%%%%%%%%%%%%%%%%%

%%%%%%%%%%%%%%%%%%%%%%%%%%%%%%%%%%%%%%%%%%%%%%%%%%%%%%%%%%
\subsection{Kinetic gravity braiding: Formalism}
%%%%%%%%%%%%%%%%%%%%%%%%%%%%%%%%%%%%%%%%%%%%%%%%%%%%%%%%%%

In this paper, we consider the KGB theory, which represents a class of scalar field theories minimally coupled to gravity. The action of the KGB theory is given as follows\footnote{We use the metric signature convention $(-+++)$ and units $G=c=1$.}
\be\label{action}
\mathcal{S}=\int d^4x\sqrt{-g}
\left[\frac{1}{16\pi}\, R +K(\phi,X) -G(\phi,X)\Box\phi\right],
\ee
where $g$ is the determinant of metric tensor $g_{\mu\nu}$, $R$ is the Ricci scalar, and $K(\phi,X)$ and $G(\phi,X)$ are arbitrary functions of the scalar field $\phi$ and the canonical kinetic term, $X=-\frac{1}{2}\nabla^\mu\phi \nabla_\mu\phi$.

Varying the action (\ref{action}) with respect to the metric $g_{\mu\nu}$, one obtains the following gravitational field equations of the KGB theory: 
\bea\label{Ein_KGB}
G_{\mu\nu}&=& 8\pi\big[K g_{\mu\nu} +K_X\nabla_\mu\phi \nabla_\nu\phi 
 - G_X\Box\phi\nabla_\mu\phi \nabla_\nu\phi 
	\nonumber \\
&&- 2\nabla_{(\mu}G \, \nabla_{\nu)}\phi
+ g_{\mu\nu}\nabla^\alpha G \, \nabla_\alpha\phi \big]. 
\eea
Furthermore, varying the action (\ref{action}) with respect to $\phi$, one obtains the equation of motion of the scalar field as follows
\be\label{EofM_KGB}
\nabla_\mu J^\mu = P,
\ee
where we have defined $J^\mu$ and $P$ as
\bea
J^\mu &=& -K_X\nabla^\mu\phi+2G_\phi\nabla^\mu\phi +G_X\Box\phi\nabla^\mu\phi +G_X \nabla^\mu X,
\Non
P &=& K_\phi +\nabla^\mu G_\phi \nabla_\mu\phi,
\non
\eea
respectively. Throughout this work, we use the notation $K_\phi=\partial K/\partial \phi$, $K_X=\partial K/\partial X$, etc.

%%%%%%%%%%%%%%%%%%%%%%%%%%%%%%%%%%%%%%%%%%%%%%%%%%%%%%%%%%
\subsection{Wormhole spacetime metric}
%%%%%%%%%%%%%%%%%%%%%%%%%%%%%%%%%%%%%%%%%%%%%%%%%%%%%%%%%%

Let us consider a static spherically symmetric wormhole configuration, given by the following spacetime metric:
\begin{equation} \label{metric}
ds^2=-A(u)dt^2+\frac{du^2}{A(u)}+r^2(u)d\Omega^2,
\end{equation}
where $d\Omega^2=d\theta^2+\sin^2\theta d\varphi^2$ is the linear element of the unit sphere, and the metric functions $A(u)$, $r(u)$, and the scalar field $\phi(u)$ are functions of the radial coordinate $u$.

To describe a traversable wormhole the metric (\ref{metric}) should possess a number of specific properties. In particular, we have: 

\begin{itemize}

\item  the radial coordinate $u$ runs through the domain $(-\infty, + \infty)$;

\item $r(u)$ has a global positive minimum at the wormhole throat $u = u_0$;

\item $A(u)$ is everywhere positive and regular, i.e., there are no event horizons and singularities in the spacetime;

\item there exist two asymptotic regions of constant curvature ${\cal R}_\pm: u\to\pm\infty$ connected by the throat.

\end{itemize}

%(i) the radial coordinate $u$ runs through the domain $(-\infty, + \infty)$; (ii) there exist two asymptotically flat regions ${\cal R}_\pm: u\to\pm\infty$ connected by the throat; (iii) $A(u)$ is everywhere positive and regular, i.e., there are no event horizons and singularities in the spacetime; (iv) $r(u)$ has a global positive minimum at the wormhole throat $u = u_0$. 

Without a significant loss of generality one can set $u_0 = 0$, so that $r_0 = \min\{r(u)\} = r(0)$ is the wormhole throat radius. Thus, the necessary conditions for the minimum of the function imposes the flaring-out condition, translated as 
\be
r_0'=0\,, \qquad  r_0'' > 0\,.  
\ee  

These condition will be used throughout this work to essentially constrain the parameters of the theory.

%%%%%%%%%%%%%%%%%%%%%%%%%%%%%%%%%%%%%%%%%%%%%%%%%%%%%%%%%%
\subsection{Field equations}
%%%%%%%%%%%%%%%%%%%%%%%%%%%%%%%%%%%%%%%%%%%%%%%%%%%%%%%%%%

Taking into account the spacetime metric (\ref{metric}), the nonzero components of the field equations (\ref{Ein_KGB}) are given by 
	\bea
	\label{KGBtt}
	2\frac{r''}{r}+\frac{r'^2}{r^2}+\frac{r'A'}{rA}-\frac{1}{r^2A} =
	\frac{8\pi K}{A}+8\pi\phi'^2 G_{\phi}
	\Non
	-4\pi A\phi'^3 G_{X}\left(\frac{A'}{A}+2\frac{\phi''}{\phi'}\right),
	\eea
	\bea
	\label{KGBuu}
	\frac{r'^2}{r^2}+\frac{r'A'}{rA}-\frac{1}{r^2A} &=&
	\frac{8\pi K}{A} +8\pi\phi'^2 K_X -8\pi\phi'^2G_{\phi}
	\Non
	&& -4\pi A\phi'^3 G_{X}\left(\frac{A'}{A}+4\frac{r'}{r}\right),
	\eea
	\bea
	\label{KGBthth}
	\frac{r''}{r}+\frac12\, \frac{A''}{A}+\frac{r'A'}{rA} =
	\frac{8\pi K}{A} +8\pi\phi'^2 G_{\phi}
	\Non	
	-4\pi A\phi'^3 G_{X}\left(\frac{A'}{A}+2\frac{\phi''}{\phi'}\right),
%	&&
%	\\
%	\label{KGBsf}
%	\frac{1}{r^2}\,\frac{d}{du}\left[
%	r^2A\phi' (K_X-2G_{\phi}) 
%	- r^2A^2\phi'^2 G_{X} \left(2\frac{r'}{r}+\frac12\frac{A'}{A}\right)\right]=
%	&&
%	\Non
%	-K_\phi -A\phi'^2 G_{\phi\phi} +\frac12 A^2\phi'^3 G_{\phi X} \left(\frac{A'}{A}+2\frac{\phi''}{\phi'}\right). 
	\eea
	respectively. 
	%where we have taken into account that 
%%
%\be
%X=-\frac{1}{2}A \phi'^2.
%\ee
%
For practical purposes, we emphasize that by combining these field equations, 
%instead of using Eqs. \Ref{KGBtt} and \Ref{KGBthth} directly, 
one can obtain a set of equations that is more compact and convenient for future analysis, given by\footnote{To obtain Eqs. (\ref{tt-uu})--(\ref{tt-thth})  we use the following combinations of equations: (\ref{KGBtt})$-$(\ref{KGBuu}), (\ref{KGBtt})$-$(\ref{KGBuu})$-2$(\ref{KGBthth}) and (\ref{KGBtt})$-$(\ref{KGBthth}), respectively.}:
\be
%(tt)-(uu)
\label{tt-uu}
\frac{r''}{r}  = 
 -4\pi\phi'^2 \bigg[K_X -2G_{\phi}    
-A\phi' G_{X}\left(2\frac{r'}{r} -\frac{\phi''}{\phi'}\right)\bigg],
%(tt)-(uu)-2(\theta\theta)
\ee
\bea
\label{tt-uu-2thth}
 (r^2A')'  =  16\pi r^2\bigg[K-XK_X +A\phi' XG_X \times
\Non   
\times \left(2\frac{r'}{r} +\frac{A'}{A}+\frac{\phi''}{\phi'}\right)\bigg],
\eea
\be
%(tt)-(\theta\theta)
\label{tt-thth}
(r^2)'' A -r^2 A'' = 2,
\ee
where the kinetic term is given by
\begin{equation}
X=-\frac{1}{2}A \phi'^2.
\end{equation}

It is worth noting that (\ref{KGBuu}) is a first order equation and hence represents a differential constraint on the values of the functions $r(u)$, $A(u)$, $\phi(u)$ and their first derivatives. The two other equations, (\ref{KGBtt}) and (\ref{KGBthth}), are of second order. 
In addition to this, one has a second order scalar field equation (\ref{EofM_KGB}), which taking into account the wormhole metric (\ref{metric}) reads
\bea
\label{KGBsf}
&&\frac{1}{r^2}\,\frac{d}{du}\left[
r^2A\phi' (K_X-2G_{\phi}) 
- \frac12 r^2A^2\phi'^2 G_{X} \left(\frac{A'}{A}+4\frac{r'}{r}\right)\right]
\Non
&&=-K_\phi -A\phi'^2 G_{\phi\phi} +\frac12 A^2\phi'^3 G_{\phi X} \left(\frac{A'}{A}+2\frac{\phi''}{\phi'}\right). 
\eea
Equation (\ref{KGBsf}) and any two second-order equations of (\ref{KGBtt})--(\ref{tt-thth}) together with the constraint (\ref{KGBuu}) represent a complete system of differential equations describing static spherically symmetric configurations in the KGB theory.

%%%%%%%%%%%%%%%%%%%%%%%%%%%%%%%%%%%%%%%%%%%%%%%
\section{Constraints at the wormhole throat}\label{sec:constraintthroat}
%%%%%%%%%%%%%%%%%%%%%%%%%%%%%%%%%%%%%%%%%%%%%%%

%{\sf *** SIMILAR ANALYSIS AS IN OUR PREVIOUS PAPER BUT $A_0'\not=0$ ***}

Let us consider the field equations at the throat $u=0$, by setting $r_0'=0$. The constraint (\ref{KGBuu}) now reads\footnote{In order not to overload the notation, here we do not use the index `0' with the functions $K$ and $G$. However, one should remember that at the throat one has $K=K(\phi_0,X_0)$, $K_X=K_X(\phi_0,X_0)$, etc.} 
\be	
\label{KGBuu0}
\frac{1}{r_0^2}=
-8\pi \left[K + A_0\phi_0'^2 \big(K_{X} - G_{\phi}\big)
-\frac12 A_0 A_0'\phi_0'^3 G_{X}\right].
\ee
This relation provides an additional constraint on the wormhole geometry. Actually, since $r_0^2 > 0$, then we obtain the condition
\be	
\label{gencond1}
K + A_0\phi_0'^2 \big(K_{X} - G_{\phi}\big)
-\frac12 A_0 A_0'\phi_0'^3 G_{X}\,<\,0.
\ee

The field equations (\ref{tt-uu})--(\ref{KGBsf}) at the throat $u=0$ can be written in the following form:
\be
\label{tt-uu0}
\frac{r_0''}{r_0} +4\pi A_0 \phi_0'^2 G_X \phi_0''=
-4\pi\phi_0'^2 \big(K_X -2G_{\phi} \big),
\ee
\be
%(tt)-(\theta\theta)
\label{tt-thth0}
\frac{r_0''}{r_0}-\frac12\,\frac{A_0''}{A_0}=\frac{1}{r_0^2A_0},
\ee
\bea
&&2 A_0^2 \phi_0'^2 G_X \frac{r_0''}{r_0} +\frac12 A_0 \phi_0'^2 G_X A_0''  
+\bigg[
-{\frac12} A_0 A_0' \phi_0'^3 G_{XX} 
	\Non 
&&-A_0 \phi_0'^2 (G_{\phi X}-K_{XX}) 
+A_0' \phi_0' G_X  +2G_\phi-K_X
\Big] A_0 \phi_0''
\Non
&& \qquad \quad = \frac14 A_0 A_0'^2 \phi_0'^4 G_{XX} 
-\frac12 A_0 A_0' \phi_0'^3 K_{XX} 
\Non
&& \qquad \qquad \qquad -\frac12 \phi_0'^2 \big(2 A_0 G_{\phi\phi}
- 2A_0 K_{\phi X} + A_0'^2 G_X\big)  
	\Non
&&  \qquad \qquad \qquad  - A_0' \phi_0' (2G_\phi-K_X) 
+K_\phi.
\eea

The above equations represent a set of linear algebraic equations for the second derivatives $r''_{0}$, $A''_{0}$ and $\phi''_{0}$. Solving with respect to $r_0''$ yields
\be\label{r2_0_gen}
\frac{r_0''}{r_0}=2\pi\phi_0'^2\,\frac{\Delta_r}{\Delta},
\ee
where for notational simplicity, we have defined the following factors
\bea
	\Delta_r &=&  
	16\pi A_0^2 A_0'\phi_0'^5 G_X^3 
	+32\pi A_0^2 \phi_0^4 G_X^2 G_{\phi}
\Non
&&
	-32\pi A_0^2 \phi_0'^4 G_X^2 K_X
	+A_0 A_0'^2 \phi_0'^4 G_X G_{XX}
\Non &&
	-2 A_0 A_0' \phi_0'^3 G_X K_{XX}
	+4 A_0 A_0' \phi_0'^3 G_{XX} G_\phi	
\Non
&&
	-2 A_0 A_0' \phi_0'^3 G_{XX} K_X
	-32\pi A_0 \phi_0'^2 G_X^2 K
\Non
&&
	-2 A_0'^2 \phi_0'^2 G_X^2
	-4 A_0 \phi_0'^2 G_X G_{\phi\phi}
\Non
&&
	+4 A_0 \phi_0'^2 G_X K_{\phi X} 
	+8 A_0 \phi_0'^2 G_\phi G_{\phi X} 
\Non
&&
	-8 A_0 \phi_0'^2 G_\phi K_{XX}
	-4 A_0 \phi_0'^2 G_{\phi X} K_X
\Non
&&
	+4 A_0 \phi_0'^2 K_X K_{XX} 
	-16 A_0' \phi_0' G_X G_\phi
\Non
&&
	+8 A_0' \phi_0' G_X K_X 
	+4 G_X K_\phi 
\Non
&&
%	-16 G_\phi^2 +16 G_\phi K_X -4 K_X^2\, ,
	-4 (2 G_\phi -K_X)^2 \, ,
\eea
and
\bea
	\Delta &=& 
	24\pi A_0^2 \phi_0'^{4} G_X^2 
	+ A_0 A_0'\phi_0'^{3} G_{XX} 
	+2 A_0\phi_0'^2 G_{\phi X} 
\Non
&&
	-2 A_0\phi_0'^2 K_{XX}  
	-2A_0'\phi_0' G_X -4G_\phi +2 K_X  ,
\eea
respectively.
Thus, the flaring-out condition $r_0''>0$ implies the following useful constraints
\be
\phi_0'^2\not=0,
\ee
and
\be
\label{gencond2}
\frac{\Delta_r}{\Delta}\,>\,0,
\ee
which should be fulfilled at the wormhole throat.

Note that it is not a trivial matter to extract much information from inequalities (\ref{gencond1}) and (\ref{gencond2}), so that in the following we will resort to specific cases.

\section{Wormhole solutions}\label{sec:specificsolution}
%%%%%%%%%%%%%%%%%%%%%%%%%%%%%%%%%%%%%%%%%%%%%%%%%%%%%%%%%%
In this section we will search for wormhole solutions in a number of specific examples of the KGB theory. 

%%%%%%%%%%%%%%%%%%%%%%%%%%%%%%%%%%%%%%%%%%%%%%%%%%%%%%%%%%
\subsection{Specific case I: $K(\phi,X)= X$ and $G(\phi,X) = f(\phi)$}
\label{case1}
%%%%%%%%%%%%%%%%%%%%%%%%%%%%%%%%%%%%%%%%%%%%%%%%%%%%%%%%%%

Let us consider the following case:
\be
K(\phi,X)= X\,, \qquad G(\phi,X) = f(\phi).
\ee

In this case, the system (\ref{tt-uu})-(\ref{tt-thth}) reduces to
%\bea
%\label{KGB_G3phi_tt}
%2\frac{r''}{r}+\frac{r'^2}{r^2}+\frac{r'A'}{rA}-\frac{1}{r^2A}&=&
%-4\pi\phi'^2(1-2f_\phi),
%\\
%\label{KGB_G3phi_rr}
%\frac{r'^2}{r^2}+\frac{r'A'}{rA}-\frac{1}{r^2A}&=&4\pi\phi'^2(1-2f_\phi),
%\\
%\label{KGB_G3phi_thetatheta}
%\frac{r''}{r}+\frac12\, \frac{A''}{A}+\frac{r'A'}{rA}&=&
%-4\pi\phi'^2(1-2f_\phi),
%\\
%\label{KGB_G3phi_eqmo}
%\left[r^2A\phi'(1-2f_\phi)\right]'&=&-r^2A\phi'^2 f_{\phi\phi}, 
%\eea   
%After some algebra, the system (\ref{KGB_G3phi_tt})-(\ref{KGB_G3phi_eqmo}) can be rewritten in the more compact and convenient form:
\bea
\label{KGB_G3phi_1}
\frac{r''}{r} &=& 4\pi\phi'^2(2f_\phi-1)\,,
\\
\label{KGB_G3phi_2}
(r^2 A')' &=& 0\,,
\\
\label{KGB_G3phi_3}
(r^2)'' A -r^2 A'' &=& 2\,.
%\\
%\label{KGB_G3phi_4}
%\left(r^2A\phi'(1-2f_\phi)\right)' &=& 0. 
\eea   
Using an appropriate parametrization, we present a solution of Eqs. (\ref{KGB_G3phi_2}) and (\ref{KGB_G3phi_3}) as follows
\be\label{KGB_metric_func}
r(u)=e^{-\mu(u)}\sqrt{u^2+a^2}, \qquad A(u)=e^{2\mu(u)},
\ee
where 
\be 
\mu(u)=(m/a)\arctan(u/a),
\ee
%
%{\color{blue}{\sf *** Maybe best to substitute the parameter $u_0$, which was previously defined as the wormhole throat, i.e., $u_0=0$, with the parameter $a$. Do you agree? ***}}\\
%{\color{red} I agree. The substitution is done.}
%
%{\color{blue}{\sf *** Can we present figures for the functions $A(u)$ and $r(u)$? ***} }
and $m$, $a$ are two free parameters. This solution represents the well-known Ellis-Bronnikov wormhole \cite{homerellis,bronikovWH} with the metric given by \cite{SusZha:08}
\be\label{EBmetric}
ds^2=-e^{2\mu}dt^2+e^{-2\mu}\left[du^2 +(u^2+a^2)d\Omega^2 \right].
\ee

\begin{figure*}[htb!]
\centering\subfigure[Profile of $r(u)$]
{\label{fig1a}\includegraphics[width=7cm]{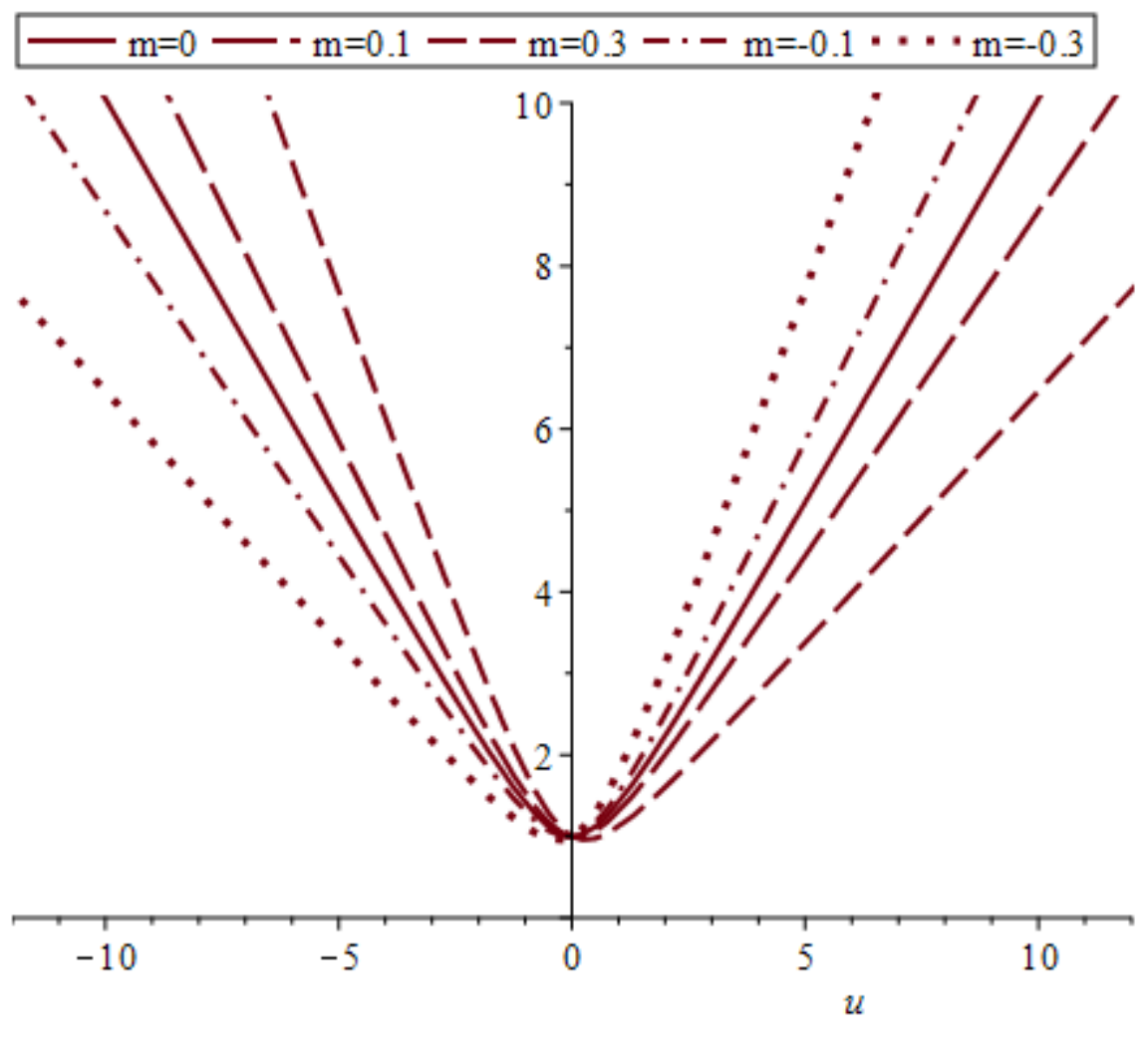} }
\hspace{0.75cm}
\subfigure[Profile of $A(u)$]
{\label{fig1b}\includegraphics[width=9.3cm]{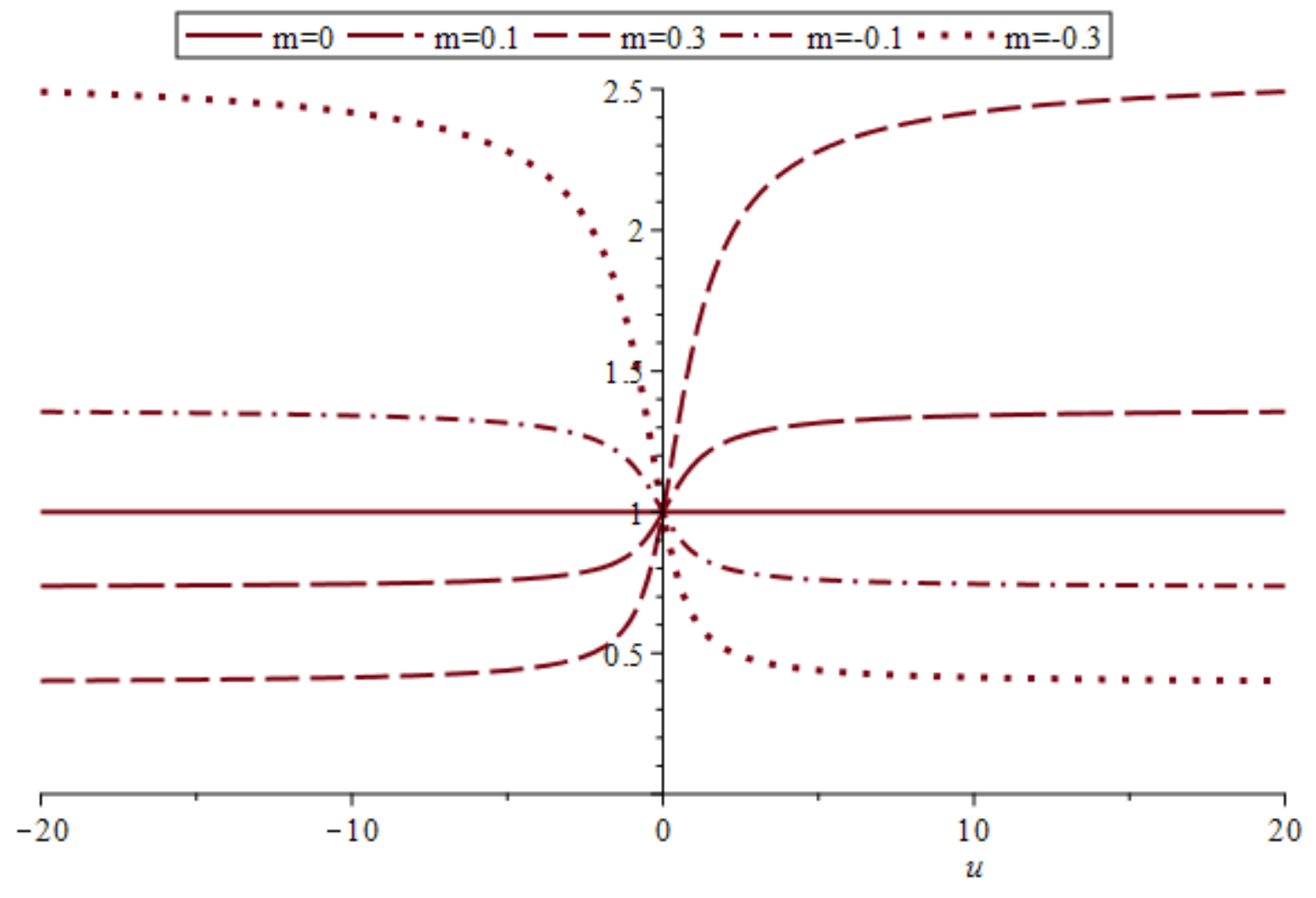} }
\caption{Profiles of $r(u)$ and $A(u)$, respectively, given by Eq. (\ref{KGB_metric_func}), for the specific case of $K(\phi,X)= X$ and $G(\phi,X) = f(\phi)$. We have considered the following values for the mass parameter, $m=0,0.1,0.3,-0.1,-0.3$. See the text for more details.}
\label{fig1}
\end{figure*}

Taking into account the following asymptotical behavior:
$$
e^{2\mu(u)}=\exp\left(\pm \frac{\pi m}{a}\right) \left[1-\frac{2m}{u}\right] +O(u^{-2}),
$$
in the limit $u\to\pm\infty$, we verify that the spacetime with the metric (\ref{EBmetric}) possesses two asymptotically flat regions where $r(u)/|u|\to {\rm const}$ and $A(u)\to \exp(\pm\pi m/a)$. These regions are connected by the throat whose radius corresponds to the minimum of the radius of the two-dimensional sphere, $r^2(u)=e^{-2\mu(u)}(u^2+a^2)$.
The graphical representations of the metric functions $r(u)$ and $A(u)$ are depicted by Fig. \ref{fig1a} and Fig. \ref{fig1b}, respectively.
   
To finalize constructing the solution, we need to find $\phi(u)$. To this effect, we substitute the expression for $r(u)$ given by (\ref{KGB_metric_func}) into Eq. (\ref{KGB_G3phi_1}) and obtain the following solution:
\be\label{KGB_phi}
\int (2f_\phi-1)^{1/2}\, d\phi = F(u)+C,
\ee
where $C$ is a constant of integration and
\be
F(u)=\left( \frac{m^2+a^2}{4\pi m^2} \right)^{1/2} \mu(u).
\ee
It is worth noticing that 
\be
\lim_{u\to\pm\infty}F(u)=F_\pm=\pm[\pi(m^2+a^2)/16a^2]^{1/2}.
\ee

Let us consider some specific examples for $f(\phi)$.

%%%%%%%%%%%%%%%%%%%%%%%%%%%%%%%%%%%%%%%%%%%%%%%%%%%%%%%%%%
\subsubsection{$f(\phi)=\frac12\lambda\phi$}
%%%%%%%%%%%%%%%%%%%%%%%%%%%%%%%%%%%%%%%%%%%%%%%%%%%%%%%%%%

In this case from Eq. (\ref{KGB_phi}) we find 
\be
\phi(u)=\frac{F(u)}{\sqrt{\lambda-1}},
\ee
provided that $\lambda>1$.

%%%%%%%%%%%%%%%%%%%%%%%%%%%%%%%%%%%%%%%%%%%%%%%%%%%%%%%%%%
\subsubsection{$f(\phi)=\frac14\lambda\phi^2$}
%%%%%%%%%%%%%%%%%%%%%%%%%%%%%%%%%%%%%%%%%%%%%%%%%%%%%%%%%%

Taking into account that
$$
\int(\lambda\phi-1)^{1/2}\,d\phi=
\frac{2}{3\lambda}(\lambda\phi-1)^{3/2},
$$
we find 
\be
\phi(u)=\frac{1}{\lambda}\left[1
+\left(\frac{3\lambda}{2}\big(F(u)+F_{+}\big)\right)^{2/3}\right],
\ee
provided that $\lambda>0$.

%%%%%%%%%%%%%%%%%%%%%%%%%%%%%%%%%%%%%%%%%%%%%%%%%%%%%%%%%%
\subsection{Specific case II:  $K(\phi,X)\equiv K(X)$ and $G(\phi,X)\equiv G(X)$}
\label{case2}
%%%%%%%%%%%%%%%%%%%%%%%%%%%%%%%%%%%%%%%%%%%%%%%%%%%%%%%%%%

In this subsection, we develop the general formalism for the specific case where the KGB factors depend solely on the kinetic term. In the following subsections we analyse particular cases of these factors.

Consider the following:
\be
K(\phi,X)\equiv K(X) \,, \qquad  G(\phi,X)\equiv G(X).
\ee

For this case the scalar field equation (\ref{KGBsf}) reduces to
	\be
	\label{KGBeqmo_X}
	\frac{d}{du}\left[
	r^2A\psi K_X 
	- \frac12 r^2A^2\psi^2 G_{X} \left(\frac{A'}{A} +4\frac{r'}{r}\right)\right] = 0,
	\ee
where we denote $\psi=\phi'$. Integrating Eq. (\ref{KGBeqmo_X}), we immediately obtain
\be\label{KGB_1int}
r^2A\psi \left[2K_X- A\psi G_X \left(\frac{A'}{A} +4\frac{r'}{r}\right)\right]  =Q,
\ee
where $Q$ is a constant of integration. Note that the first integral (\ref{KGB_1int}) represents an additional constraint on the values of the functions $r(u)$, $A(u)$, $\phi(u)$ and their first derivatives.
Using this first integral, we rewrite the constraint (\ref{KGBuu}) as follows
\be\label{uu_gx}
\frac{r'^2}{r^2}+\frac{r'A'}{rA}-\frac{1+4\pi Q\psi}{r^2A}=\frac{8\pi K}{A}.
\ee
Furthermore, substituting Eq. (\ref{KGB_1int}) into (\ref{tt-uu}), we find
\be\label{tt-uu_gx}
\frac{r''}{r}=-2\pi\psi^2\left[\frac{Q}{r^2A\psi} +A\psi G_X \left(\frac{A'}{A} +2\frac{\psi'}{\psi}\right)\right].
\ee

The set of equations (\ref{tt-thth}), (\ref{KGBeqmo_X}) and (\ref{tt-uu_gx}) together with the constraints (\ref{KGB_1int}) and (\ref{uu_gx}) form a complete system of ordinary second-order differential equations for the functions $r(u)$, $A(u)$ and $\psi(u)$ (shortly `the system'). 
%describing static spherically symmetric configurations in the KGB theory \Ref{action} with $K=K(X)$ and $G=G(X)$. 
The initial conditions for the system are fixed at the wormhole throat $u_0=0$ as $r_0$, $r_0'$, $A_0$, $A_0'$ and $\psi_0$, where $r(0)=r_0$, etc. Note that the flaring-out condition imposes $r_0'=0$. Moreover, using an appropriate rescaling of the time coordinate $t$, one can set $A_0=1$. The remaining three quantities $r_0$, $A_0'$ and $\phi_0$ are connected by two constraints, (\ref{KGB_1int}) and (\ref{uu_gx}). 

At the wormhole throat these constraints can be recast as follows
%Eq. \Ref{uu_gx} at the throat reads
%
\be\label{KGB_r0}
\frac{1}{r_0^2}=-\frac{8\pi K}{1+4\pi Q\psi_0},
\ee
%
%Demanding the positivity of the right hand side of the above relation gives
%%
%\be
%\frac{8\pi K}{1+4\pi Q\psi_0} < 0.
%\ee
%
%The first integral (\ref{KGB_1int}) can be now recast as follows 
%
and
\be\label{KGB_1int_0}
A_0'=\frac{2A_0\psi_0 K_X(1+4\pi Q\psi_0) +8\pi QK}{\psi_0^2 G_X(1+4\pi Q\psi_0)}.
\ee
Therefore, the only free parameter of the initial conditions is $\psi_0$.

Solving the system with respect to the higher derivatives $r''$, $A''$, and $\psi'$ and taking into account the relations (\ref{KGB_r0}) and (\ref{KGB_1int_0}), we obtain the following expression:
\be\label{r2_0_gx}
\frac{r_0''}{r_0}=4\pi\psi_0\,\frac{\Delta_r}{\Delta},
\ee
where
\bea
\Delta_r &=& -2 \pi \psi_0^4 G_X^3 K 
+\frac{1}{4}\psi_0^2 G_X K_X^2
\Non
&&
+ \pi Q\psi_0^3 \big[-8\pi\psi_0^2 G_X^3 K
+ G_{XX} K_X K 
\Non
&&
+2 G_X K_X^2 -G_X K_{XX} K \big]
\Non
&&
+4 \pi^2 Q^2 \big[ \psi_0^4 (G_{XX} K_X K  +G_X K_X^2 
\Non
&&
-G_X K_{XX} K ) +\psi_0^2 G_{XX} K^2 
\Non
&&
-2 G_X K^2\big],
\eea
and
\bea
\Delta &=& 
(1+4\pi Q\psi_0 ) 
\Big[3\pi\psi_0^5 G_X^3
\Non
&&
+\frac14\psi_0^3(G_{XX} K_X -G_X K_{XX})
%\Non
%&&
-\frac14\psi_0 G_X K_X 
\Non
&&
+\pi Q\, \big[12\pi\psi_0^6 G_X^3
%\Non
%&&
+\psi_0^4 (G_{XX} K_X -G_X K_{XX})
\Non
&&
+\psi_0^2 (G_{XX} K -G_X K_X)
%\Non
%&&
-2 G_X K\big]
\Big].
\eea
The flaring-out condition, $r_0''>0$, imposes the following inequality
\be
\label{cond1_gx}
\psi_0\frac{\Delta_r}{\Delta}\,>\,0,
\ee
which should be fulfilled at the wormhole throat. The above inequality imposes restrictions for the values of $\psi_0$ and, as a consequence, $r_0$ and $A_0'$. An additional restriction on the initial value $\psi_0$ follows from Eq. (\ref{KGB_r0}). Demanding the positivity of the right hand side of Eq. (\ref{KGB_r0}) yields
\be\label{cond2_gx}
\frac{8\pi K}{1+4\pi Q\psi_0} < 0.
\ee

The system of field equations (\ref{tt-thth}), (\ref{KGBeqmo_X}) and (\ref{tt-uu_gx}) supplemented by the conditions (\ref{KGB_r0}), (\ref{KGB_1int_0}), (\ref{cond1_gx}) and (\ref{cond2_gx}) can be analyzed numerically. To this effect, we further consider several specific examples below.

\subsection{Specific case III:  $K(X)=X$ and $G(X)=\lambda X$}

\begin{figure*}[htb!]
\centering\subfigure[Behaviour of $r(u)$]
{\label{fig2a}\includegraphics[width=7cm]{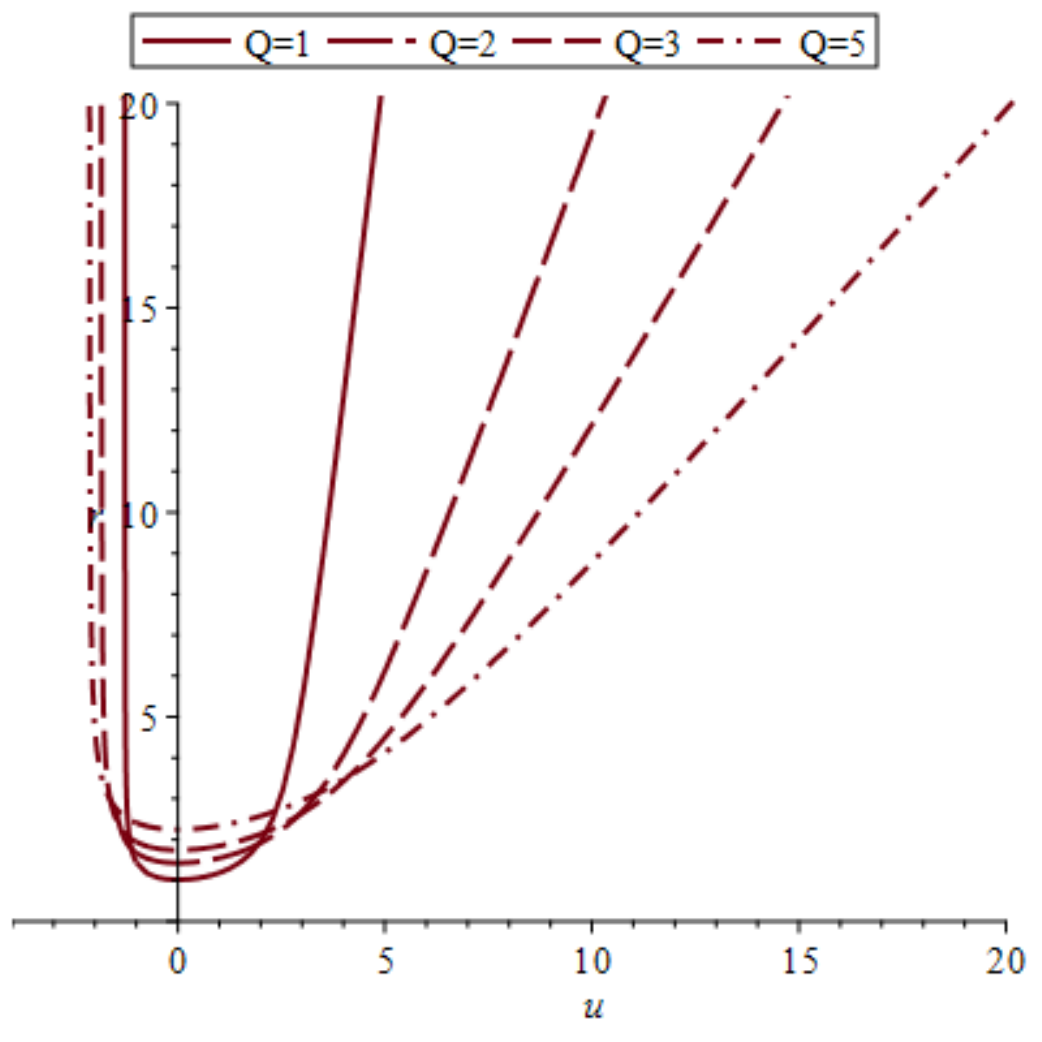} }
\hspace{0.75cm}
\subfigure[Behaviour of $A(u)$]
{\label{fig2b}\includegraphics[width=7cm]{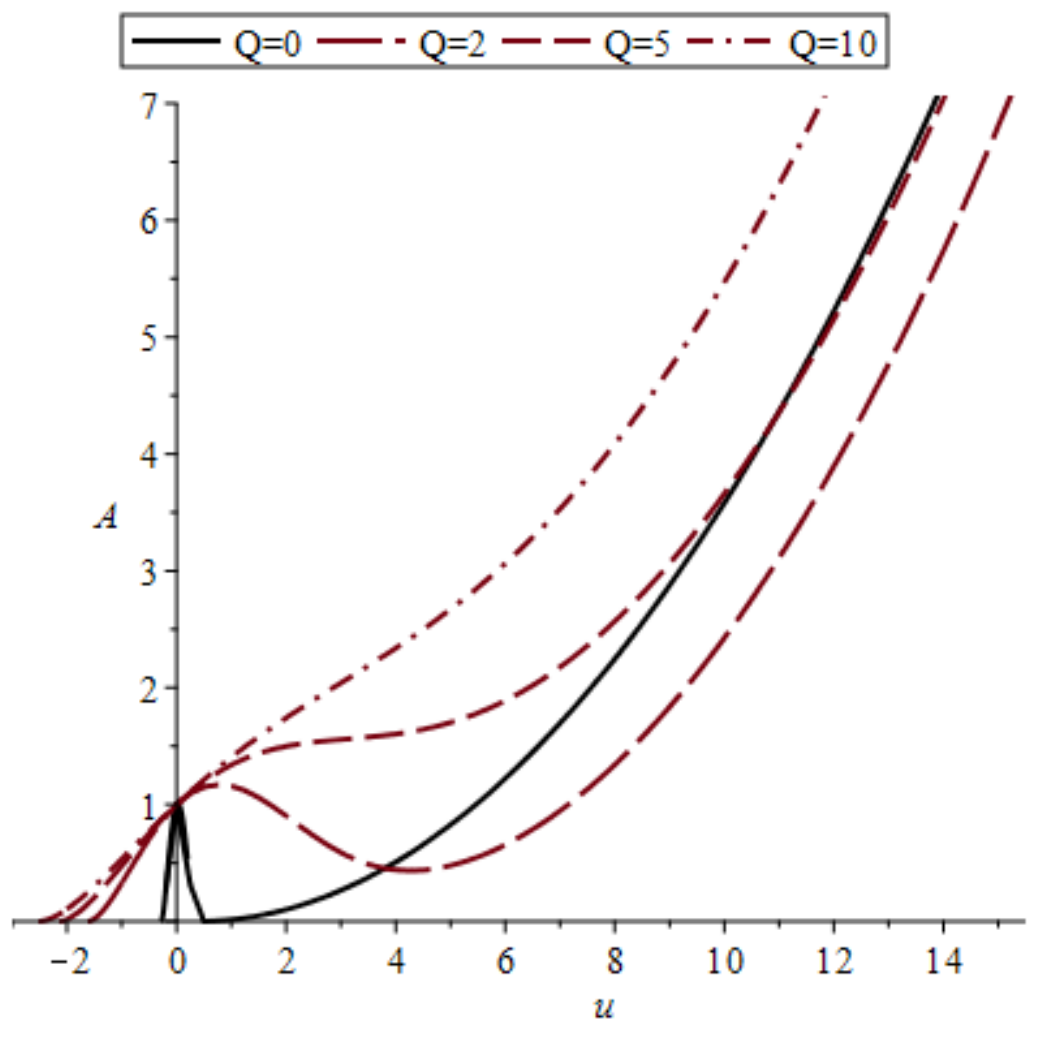} }
\caption{Behaviour of $r(u)$ and $A(u)$ in the vicinity of the wormhole throat for the model with $K(X)=X$ and $G(X)=\lambda X$. For \ref{fig2a}, the curves are plotted for $\lambda=1$ and $Q=1,2,3,5$; while for \ref{fig2b}, the curves are given for $\lambda=1$ and $Q=0,2,5,10$. See the text for more details.}
\label{fig_(K=X_G=phi_m=0_01_03)}
\end{figure*}

Using the set of equations of the previous subsection, here we present the results of the numerical analysis obtained for the KGB model (\ref{action}) with $K(X)=X$ and $G(X)=\lambda X$. 

A typical solution for the radial metric function $r(u)$ is shown in Fig. \ref{fig2a}.
It is seen that $r(u)$ has the global minimum, so that a throat exists at $u=0$. Far away from the throat, in the region $u\gg 0$, $r(u)$ behaves linearly as $r(u)\approx \alpha u$, where $\alpha=\lim_{u\to\infty}[r(u)/u]$. On the other side of the throat, in the region $u<0$, $r(u)$ goes to infinity at $u=u_*$, where $u_*<0$ is a numerical parameter which depends on the initial conditions. The numerical analysis gives $r^2(u)\approx 1/[\alpha_*(u-u_*)]$ at $u\to u_*$.
%In Fig. \ref{figr} the graph of $r(u)$ is shown for $A_0'=0.1$; correspondingly, $\alpha=0.16$ and $u_*=-10.087$.   

A typical behavior of the metric function $A(u)$ 
%far from the wormhole throat 
is shown in Fig. \ref{fig2b}.
Far away from the throat, in the asymptotic region $u \gg 0$ the function $A(u)$ has the asymptotics $A(u)\approx(\beta u)^2$, where $\beta^2=\lim_{u\to\infty}[A(u)/u^2]$. Therefore, at $u \gg 0$, the spacetime metric takes the following asymptotic form
\be
ds^2 \approx -(\beta u)^2 dt^2 +\frac{du^2}{(\beta u)^2} +(\alpha u)^2 d\Omega^2,
\ee
and the corresponding scalar curvature is $R=-12\beta^2 +O\big((\alpha u)^{-2}\big)$. Therefore, the spacetime far from the throat is asymptotically anti-de Sitter with a constant negative curvature $R=-12\beta^2$. Numerical values of the parameters $\alpha$ and $\beta$ depend on the initial conditions at the throat. 

In the vicinity of the throat, the behaviour of $A(u)$ is more varied and depends on the value of the constant of integration $Q$. For $Q>Q_*\approx 4.5$ the function $A(u)$ increases monotonically within the interval $(u_*,\infty)$. In the case $Q<Q_*$ the function $A(u)$ is not monotonic. It first increases monotonically within the interval $(u_*,u_1)$ and achieves its local maximum at $u=u_{1}>0$. Then, $A(u)$ decreases and reaches a local minimum at $u=u_2>u_1$. 

In the region $u<0$, $A(u)$ decreases and goes to zero at $u=u_*$. The numerical analysis reveals the following asymptotic behavior near $u=u_*$: $A(u)\approx \beta_*^2 (u-u_*)^2$. Therefore, the spacetime metric can be approximately written as
\be\label{near_u_asterix}
ds^2 \approx -\beta_*^2(u-u_*)^2 dt^2 +\frac{du^2}{\beta_*^2(u-u_*)^2} +\frac{ d\Omega^2}{\alpha_*(u-u_*)}.
\ee  
One can see that the metric (\ref{near_u_asterix}) has an explicit singular feature near $u=u_*$. However, this may be just a coordinate artifact. For example, introducing the new radial coordinate $\tilde u=\ln(u-u_*)$, one can remove this singularity. In order to characterize adequately the spacetime geometry one must analyze curvature invariants. In particular, the scalar curvature corresponding to the metric (\ref{near_u_asterix}) is $R=-\frac32\beta_*^2 +O\big(\alpha_*(u-u_*)\big)$. 
Therefore, the spacetime in the region near $u=u_*$ is asymptotically anti-de Sitter with a constant negative curvature $R=-\frac32\beta_*^2$. 

Summarizing the results obtained in this subsection, we conclude that the KGB theory (\ref{action}) with model functions $K(X)=X$ and $G(X)=\lambda X$ admits solutions describing static wormholes whose throats connect two anti-de Sitter spacetimes.

\subsection{Specific case IV:  $K(X)=X +\mu^2 X^2$ and $G(X)=\lambda X$}

\begin{figure*}[htb!]
\centering\subfigure[Behaviour of $r(u)$]
{\label{fig3a}\includegraphics[width=7cm]{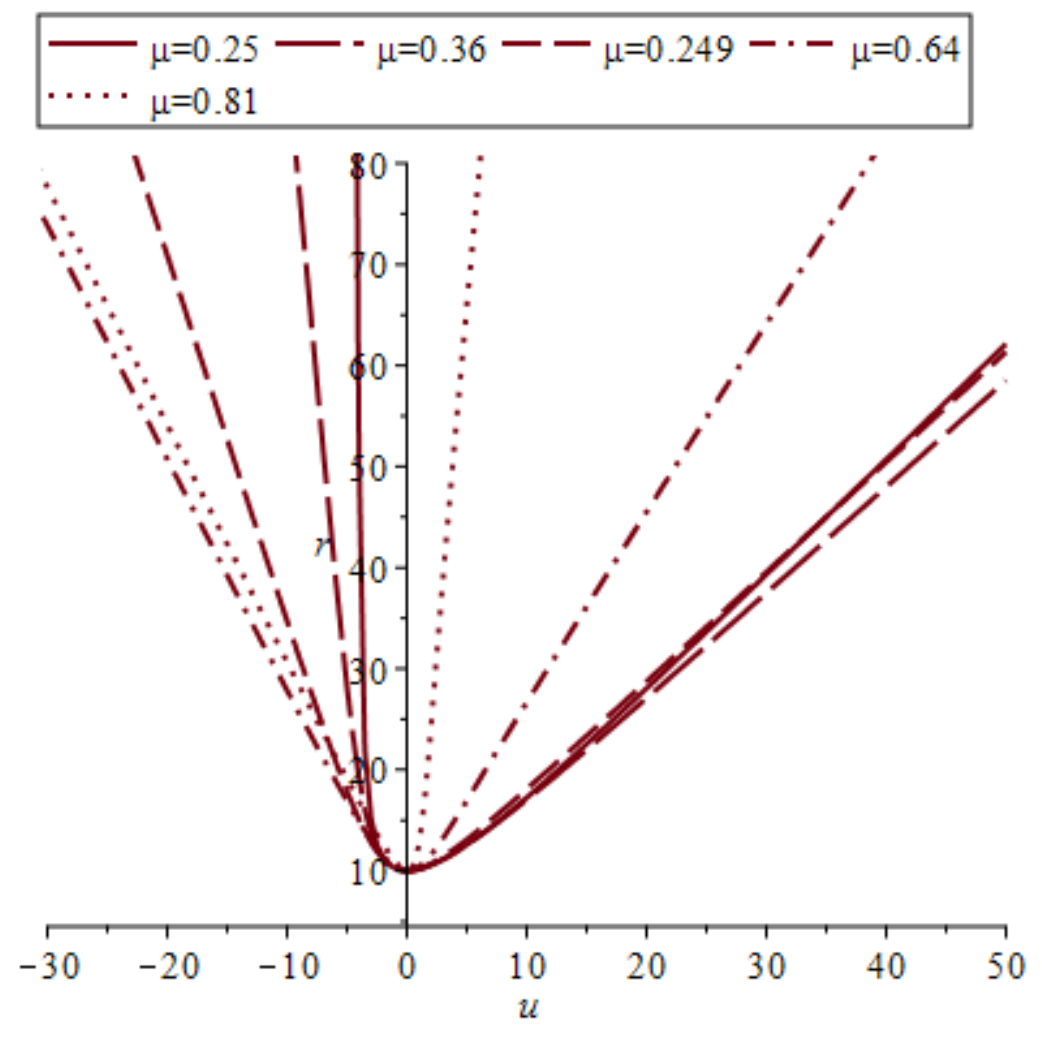} }
\hspace{0.75cm}
\subfigure[Behaviour of $A(u)$]
{\label{fig3b}\includegraphics[width=7cm]{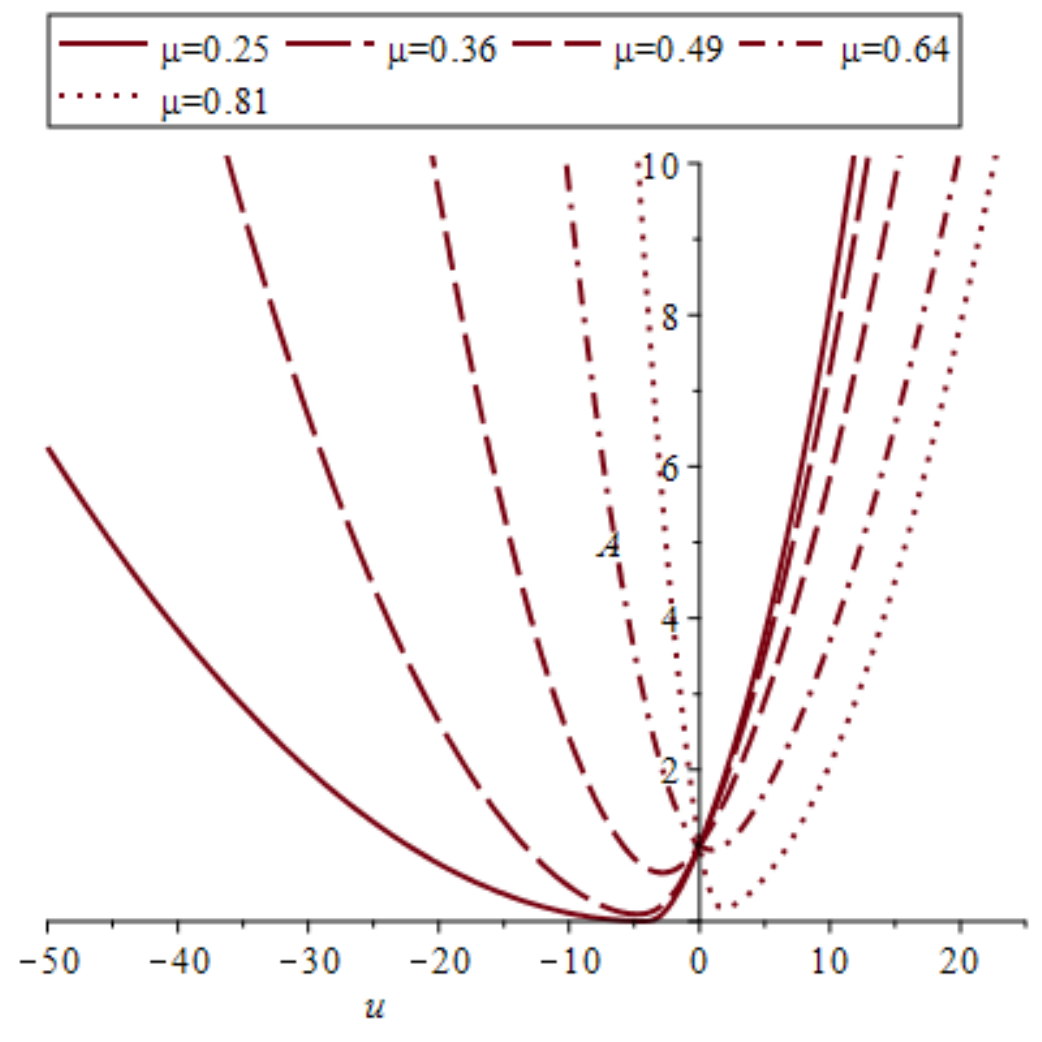} }
\caption{Behaviour of $r(u)$ and $A(u)$ in the vicinity of the wormhole throat for the model with $K(X)=X+\mu^2 X^2$ and $G(X)=\lambda X$. The curves are plotted for $\lambda=1$, $\mu=0.25, 0.36, 0.49, 0.64, 0.81$ and $Q=100$. See the text for more details.}
\label{fig3}
\end{figure*}
\begin{figure*}[htb!]
\centering\subfigure[Behaviour of $A(u)$ for $\mu=0.36$ and $Q=1, 3, 5, 10$]
{\label{fig3c}\includegraphics[width=7cm]{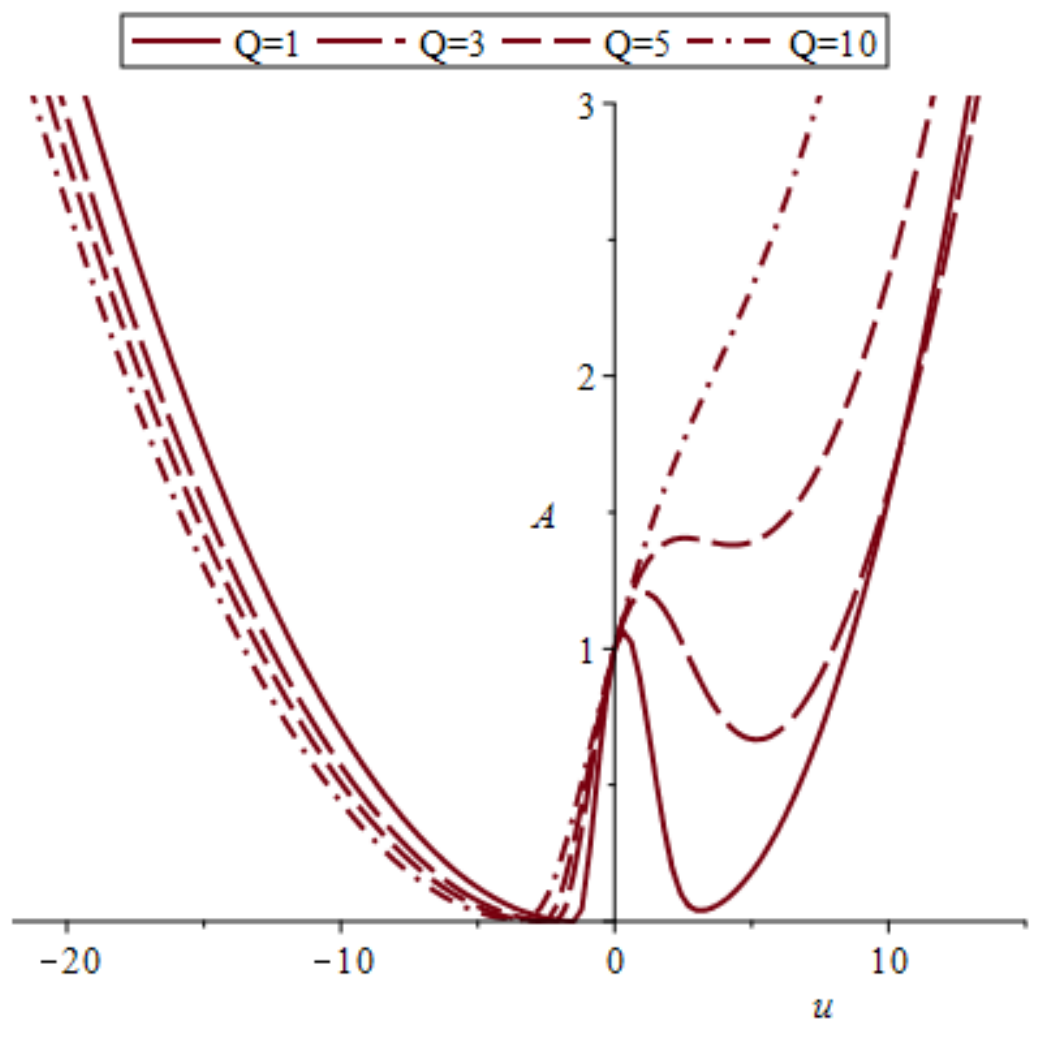} }
\hspace{0.75cm}
\subfigure[Behaviour of $A(u)$ for $\mu=0.81$ and $Q=0.1, 0.3, 0.5, 1, 5$]
{\label{fig3d}\includegraphics[width=7cm]{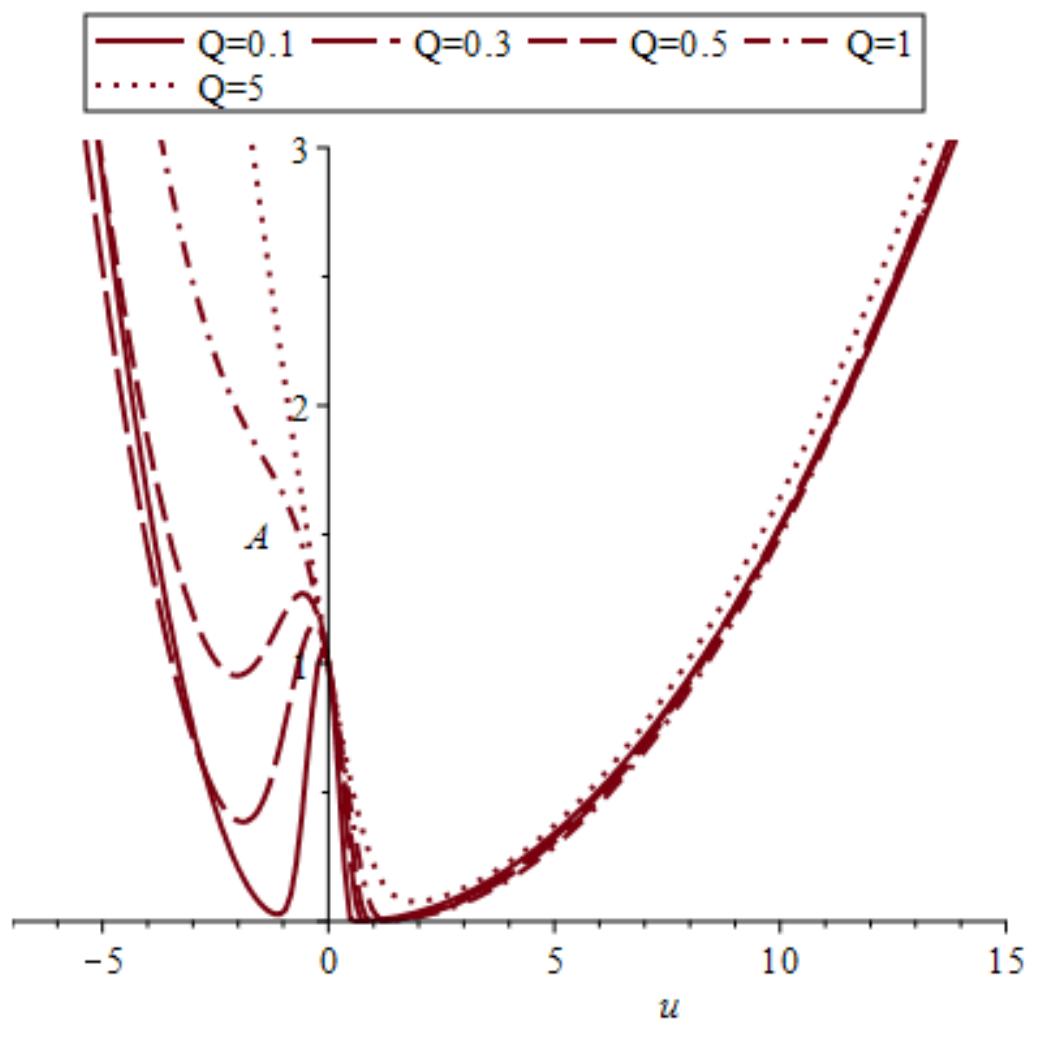} }
\caption{Behaviour of $A(u)$ in the vicinity of the wormhole throat in the model with $K(X)=X+\mu^2 X^2$ and $G(X)=\lambda X$. In Fig. \ref{fig3c}, the curves are plotted for $\lambda=1$, $\mu=0.36$ and $Q=1, 3, 5, 10$; in Fig. \ref{fig3d}, for $\lambda=1$, $\mu=0.81$ and $Q=0.1, 0.3, 0.5, 1, 5$. See the text for more details.}
\label{fig3_2}
\end{figure*}

In this subsection we consider the KGB theory (\ref{action}) with the model function $K(X)$ being quadratic with respect to $X$. For this aim, we assume that $K(X)=X +\mu^2 X^2$ and $G(X)=\lambda X$. The results of the numerical analysis are presented in Fig. \ref{fig3}.

More specifically, the radial metric function $r(u)$ given in Fig. \ref{fig3a} has the global minimum $r_0$ at the throat $u=0$. Far from the throat, in the asymptotic regions  ${\cal R_\pm}: u\to\pm\infty$, one has $r(u)=\pm\alpha_\pm |u|+O(1/|u|)$.
%, where $\alpha_\pm$ are parameters which depend on initial values. 

The typical behavior of the metric function $A(u)$ is shown in Fig. \ref{fig3b}. It takes a minimal value at the region near the wormhole throat $u=0$. In the asymptotic regions  ${\cal R_\pm}: u\to\pm\infty$, one has $A(u)=\beta_\pm^2 u^2+O(1/|u|)$.
%, where $\beta_\pm$ are parameters which depend on initial values.

The asymptotic form of the spacetime metric in the regions ${\cal R_\pm}: u\to\pm\infty$ reads
\be
ds_\pm^2 \approx -(\beta_\pm u)^2 dt^2 +\frac{du^2}{(\beta_\pm u)^2} +(\alpha_\pm u)^2 d\Omega^2.
\ee
This metric corresponds to an anti-de Sitter spacetime with a constant negative curvature $R=-12\beta_\pm^2$. Numerical values of the parameters $\alpha_\pm$ and $\beta_\pm$ depend on the initial conditions at the throat. 

It is also worth noting that the behavior of $A(u)$ in the vicinity of the throat depends essentially on the value of the integration parameter $Q$. This behavior is illustrated in Figs. \ref{fig3c} and \ref{fig3d}. It is seen that for relatively small values of $Q$ the metric function $A(u)$ possesses a local maximum, that is, there exists a gravitational ``barrier'' near the wormhole throat. The height of this barrier decreases for increasing values of $Q$, and finally disappears for large values of $Q$.

Thus, we conclude that the KGB theory (\ref{action}) with the functions $K(X)=X +\mu^2 X^2$ and $G(X)=\lambda X$ provides solutions describing static wormholes whose throats connect two anti-de Sitter spacetimes.

%%%%%%%%%%%%%%%%%%%%%%%%%%%%%%%%%%%%%%%%%%%%%%%%%%%%%%%%%%
\section{Tidal forces}
%%%%%%%%%%%%%%%%%%%%%%%%%%%%%%%%%%%%%%%%%%%%%%%%%%%%%%%%%%

An interesting feature of wormholes is their traversability, namely, one may consider whether the conditions imposed imply that the tidal forces are small enough to assure that the throat can be safely crossed. To this effect, note that the tidal effect is proportional to the Riemann tensor (we refer the reader to Eq. (13.1), Chapter 13, in \cite{Visser:1995cc} for specific details):
\be
(\Delta a)^\mu=-R^\mu_{\alpha\nu\beta}V^\alpha (\Delta\xi)^\nu V^\beta.
\ee

The radial and transverse components are given by the following
\bea
(\Delta a)_\parallel &=& -R_{\hat t \hat u \hat t \hat u} (\Delta\xi)_\parallel,
%\ee
%\be
\\
(\Delta a)_\perp &=&-\gamma^2 \left(R_{\hat\theta \hat t \hat\theta \hat t}+
\beta^2 R_{\hat\theta \hat u \hat\theta \hat u}\right) (\Delta\xi)_\perp,
\eea	
where $\beta=v/c$ and $\gamma=1/\sqrt{1-\beta^2}$. The components of the Riemann tensor are given in terms of metric (\ref{metric}), and we consider an orthonormal frame denoted by the hats on the indices. Thus, in terms of the metric functions $A(u)$ and $r(u)$ one has
\bea
(\Delta a)_\parallel &=& -\frac12 A'' (\Delta\xi)_\parallel,
\\
(\Delta a)_\perp &=& \left[\frac12 A'\frac{r'}{r} 
-\beta^2\gamma^2 A\frac{r''}{r}\right] (\Delta\xi)_\perp \,.
\eea
If an object of size $l$ is designed to withstand a maximum tidal acceleration $g$ and is to traverse the wormhole safely, one must have
\bea
\frac12\left|A''\right| &\le& \frac{g}{l},
\\
\left|\frac12 A'\frac{r'}{r} 
-\beta^2\gamma^2 A\frac{r''}{r}\right| &\le& \frac{g}{l} \,.
\eea
At the throat the tidal inequalities yield
\bea
\frac12\left|A_0''\right| &\le& \frac{g}{l},
\\
\beta^2\gamma^2 \left| \frac{r_0''}{r_0}\right| &\le& \frac{g}{l}\,,
\eea
respectively.

In the following, we consider the solutions provided in subsections \ref{case1} and \ref{case2}, in order to illustrate the traversability conditions imposed on these specific wormhole configurations.

%%%%%%%%%%%
\subsection{Tidal forces in case I: $K(\phi,X)= X$ and $G(\phi,X) = f(\phi)$}
%%%%%%%%%%%

In this case the wormhole solution is given by the explicit form (\ref{KGB_metric_func}). Using this solution, we find the following tidal inequalities
\bea
\frac{2m^2}{a^4} &\le& \frac{g}{l},
\\
\beta^2\gamma^2 \frac{m^2+a^2}{a^4} &\le& \frac{g}{l},
\eea
which place stringent restrictions for the wormhole parameters $m$ and $a$, as for the transit velocity $\beta$.

%%%%%%%%%%%
\subsection{Tidal forces in the case II: $K(\phi,X)= K(X)$ and $G(\phi,X) = G(X)$}
%%%%%%%%%%%

Using Eqs. (\ref{tt-thth0}), (\ref{KGB_r0}) and (\ref{r2_0_gx}), we can rewrite the tidal inequalities as follows
\bea
4\pi\left|\psi_0\frac{\Delta_r}{\Delta}+\frac{2K}{1+4\pi Q\psi_0}\right| &\le& \frac{g}{l},
\\
4\pi \beta^2\gamma^2 \left| \psi_0\frac{\Delta_r}{\Delta} \right| &\le& \frac{g}{l}.
\eea
These inequalities impose the restrictions for the values of $\psi_0$ and $\beta$ which provide the the traversability conditions through the wormhole throat.

%%%%%%%%%%%%%%%%%%%%%%%%%%%%%%%%%%%%%%%%%%%%%%%%%%%%%%%%%%
\section{Conclusions}\label{sec:conclusions}\label{sec:conclusion}
%%%%%%%%%%%%%%%%%%%%%%%%%%%%%%%%%%%%%%%%%%%%%%%%%%%%%%%%%%

In this work, we have continued the research project initiated in Ref. \cite{Korolev:2020ohi}, where we explored a wide variety of subclasses of the Horndeski Lagrangian models, namely, quintessence/phantom fields, $k$-essence, scalar-tensor theories, covariant galileons, nonminimal kinetic coupling, kinetic gravity braiding, and the scalar-tensor representation of Gauss-Bonnet couplings, amongst others, and considered the analysis restricted to the wormhole throat. In fact, this proved to be an extremely useful study, as it served as a consistency check for a plethora of solutions obtained in the literature and indeed paved the way for new avenues of research related to subclasses of Horndeski wormhole solutions.

Thus, in this work, following the strategy outlined above, we extended the analysis of Ref. \cite{Korolev:2020ohi} and considered the possibility that traversable wormhole geometries were sustained within the kinetic gravity braiding theory. In this context, we presented the full gravitational field equations in a static and spherically symmetric traversable wormhole background, and using the  flaring-out conditions we deduced the general constraints at the wormhole throat. Using this analysis, we then presented a plethora of wormhole solutions using analytical and numerical methods by considering particular choices of the KGB factors. 
An interesting feature of these solutions related to the energy conditions, consists in that we are essentially considering a scalar-tensor theory of gravity without any ordinary matter. In this case the wormhole solutions obtained are, in a certain sense, vacuum solutions of the theory and thus consequently do not require exotic matter.
In conclusion, the analysis carried out explicitly demonstrated that the KGB theory exhibits a rich structure of wormhole geometries, ranging from asymptotically flat solutions to asymptotically anti-de Sitter spacetimes.

%%%%%%%%%%%%%%%%%%%%%%%%%%%%%%%%%%%%%%%%%%%%%%%%%%%%%%%%%%
\section*{Acknowledgments}
FSNL acknowledges support from the Funda\c{c}\~{a}o para a Ci\^{e}ncia e a Tecnologia (FCT) Scientific Employment Stimulus contract with reference CEECIND/04057/2017. FSNL also thanks funding from the research grants No. UID/FIS/04434/2020, No. PTDC/FIS-OUT/29048/2017 and No. CERN/FIS-PAR/0037/2019.
S.V.S. and R.K. are supported by the RSF grant No. 16-12-10401. Partially, this work was done in the framework of the Russian Government Program of Competitive Growth of the Kazan Federal University.
%%%%%%%%%%%%%%%%%%%%%%%%%%%%%%%%%%%%%%%%%%%%%%%%%%%%%%%%%%

%%%%%%%%%%%%%%%%%%%%%%%%%%%%%%%%%%%%%%%%%%%%%%%%%%%%%%%%%%

%%%%%%%%%%%%%%%%%%%%%%%%%%%%%%%%%%%%%%%%%%%%%%%%%%%%%%%%%%

%%%%%%%%%%%%%%%%%%%%%%%%%%%%%%%%%%%%%%%%%%%%%%%%%%%%%%%%%%
\end{document}